\documentclass[aps,twocolumn,prd,floats,nofootinbib,showpacs]{revtex4}

\usepackage{graphicx}
\usepackage{amsmath}
\usepackage{amsfonts}
\usepackage{amssymb}

\def\pbi{~\mbox{pb}^{-1}}

\def\mh{m_h}

\def\hsm{h_{SM}}
\def\mhsm{m_{\hsm}}

\def\beq{\begin{equation}}
\def\eeq{\end{equation}}
\def\bea{\begin{eqnarray}}
\def\eea{\end{eqnarray}}

\def\lsim{\mathrel{\raise.3ex\hbox{$<$\kern-.75em\lower1ex\hbox{$\sim$}}}}
\def\gsim{\mathrel{\raise.3ex\hbox{$>$\kern-.75em\lower1ex\hbox{$\sim$}}}}
\def\ifmath#1{\relax\ifmmode #1\else $#1$\fi}

\def\br{BR}
\def\gev{~{\mbox{GeV}}}

\def\to{\rightarrow}

\def\hsm{h_{SM}}

\def\lam{\lambda}

\def\anti{\overline}

    \def\fillboxx#1#2{\hbox to #1{\vbox to #2{\vfil}\hfil}   }

\def\tauptaum{\tau^+\tau^-}

\def\gev{~{\rm GeV}}
\def\gam{\gamma}
\def\tanb{\tan\beta}
\def\cotb{\cot\beta}

\def\cosb{\cos\beta}
\def\sinb{\sin\beta}
\def\cosa{\cos\alpha}
\def\sina{\sin\alpha}

\def\anti{\overline}

\def\mgamgam{M_{\gam\gam}}
\def\rgamgam{R_{\gam\gam}}
\def\rgamgama{\rgamgam^A}
\def\rgamgamh{\rgamgam^h}
\def\rgamgamH{\rgamgam^H}
\def\rww{R_{WW}}
\def\rzz{R_{ZZ}}

\def\rwwh{\rww^h}

\def\gamgg{\Gamma_{gg}}

\def\brpp{\br(\gam\gam)}

\newcommand{ \slashchar }[1]{\setbox0=\hbox{$#1$}   
   \dimen0=\wd0                                     
   \setbox1=\hbox{/} \dimen1=\wd1                   
   \ifdim\dimen0>\dimen1                            
      \rlap{\hbox to \dimen0{\hfil/\hfil}}          
      #1                                            
   \else                                            
      \rlap{\hbox to \dimen1{\hfil$#1$\hfil}}       
      /                                             
   \fi}     
\begin{document}
\title{\boldmath Ruling out a fourth generation using limits on hadron
  collider Higgs 
  signals 
}

\author{John  F. Gunion}

\affiliation{ Department of Physics, University of California, Davis,
  CA 95616, USA}

\begin{abstract}
  We consider the impact of a 4th generation on Higgs to $\gam\gam$
  and $WW,ZZ$ signals and demonstrate that the Tevatron and LHC have
  essentially eliminated the possibility of a 4th generation if the
  Higgs is SM-like and has mass below $200\gev$.  We also show that
  the absence of enhanced Higgs signals in current data sets in the $\gam\gam$
  and $WW,ZZ$ final states can strongly constrain 
  the possibility of a 4th generation in 
  two-Higgs-doublet models of type II, including the MSSM.

\end{abstract}


\maketitle

Although new physics has not yet been seen at the Tevatron or LHC, as
the integrated luminosity, $L$, escalates increasingly interesting
constraints on new physics emerge. This Letter focuses on the
interconnection between limits on excesses in the $\gam\gam$ and
$WW,ZZ$ mass spectra and the possible existence of a 4th generation
and/or a sequential $W'$, assuming existence of: (1) a Standard Model
(SM) Higgs boson; or (2) a two-doublet Higgs sector (including the
special case of the Minimal Supersymmetric Standard Model,
MSSM). Important results arise even though a Higgs boson has not yet
been detected.

There are now significant constraints on Higgs to $\gam\gam$ and $WW$
signals coming from the current Tevatron and LHC data samples. A
convenient review is Ref.~\cite{atlaspheno}.  In particular, no peak
is observable in the $\gam\gam$ channel in the $L=131\pbi$ ATLAS data,
and, indeed, the observed rate lies somewhat below the expected
background rate. Similarly, both the LHC and, especially, the Tevatron
restrict any excess in the $WW$ channel relative to the SM rate.  We
define the ratio $R_X^h\equiv [\Gamma_{gg}^h\br(h\to
X)]/[\Gamma_{gg}^{\hsm}\br(\hsm\to X)]$, where the denominator is
always computed for 3 generations. Crude estimates from the ATLAS
$\gam\gam$ spectrum plots of \cite{atlaspheno} are that $\rgamgam\lsim
10$ for $\mgamgam$ in the $100-150\gev$ range.  As regards $\rww$,
currently the Tevatron CDF+D0 combination~\cite{:2010ar} provides the
strongest limits: at 95\% CL the Bayesian upper limits on $\rww$ in
the $m_h\in[100,200]\gev$ window range between 2.54 and 0.64. 
Limits of this same order will eventually be achieved
out to large $m_{WW}$ as $L$ increases.  

These constraints motivate an
examination of the possibilities for enhanced $\rgamgam$ and $\rww$
values in the context of various models for the Higgs sector.  Here,
we consider implications for a 4th generation in the context of the
Standard Model (SM) and two-Higgs-doublet models (2HDM) (including the
MSSM) and for a sequential $W'$ in the SM case.  The lepton and quark
masses of the 4th generation will be set to $400\gev$ and $1400\gev$
will be chosen for the $W'$ mass, both only slightly above current
experimental limits.

\begin{figure}
\vspace*{-.4in}
\includegraphics[height=0.5\textwidth,angle=90]{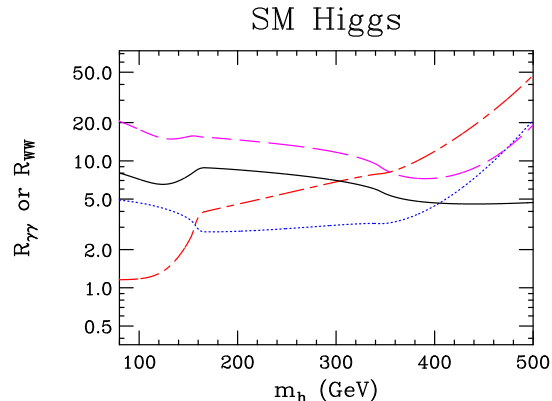}
\vspace*{-.5in}
\caption{The solid black curve shows $\rww$ in the presence of a 4th
  generation.  For $\rgamgam$: the long-dash -- short-dash red curve
  is for a 4th generation only; the dotted blue curve is for a
  sequential $W'$ only; the long-dash magenta curve is for a 4th
  generation plus a sequential $W'$.  All curves are for a Higgs boson
  with SM-like couplings and SM final decay states.\vspace*{-.2in}}
\label{htohsm4}
\end{figure}
A plot showing $\rgamgam$ and $\rww$ as a function of $m_h$ in the
case of an $h$ with SM-like couplings and decays appears in
Fig.~\ref{htohsm4}. (See also \cite{Ruan:2011qg}.) If a 4th generation
is present, one observes large $\rgamgam$ ($\geq 4$) only for
$m_h>2m_W$,\footnote{$\rgamgam\sim 1$ for $m_h\lsim 130\gev$ because
  the increase in $\gamgg$ is closely offset by a decrease in $\brpp$
  resulting from the increased cancellation of the 4th generation
  fermion loops with the (opposite sign) $W$ loop.} where, in any
case, prospects for probing $\rgamgam\leq 4$ must be regarded as
uncertain due to the large size of the Higgs total width.
Fortunately, the $WW$ channel is much more definitive. $\rww$, also
plotted in Fig.~\ref{htohsm4}, is predicted to be $\geq 6.5$ for
$m_h<300\gev$, falling to $\geq 4.8$ for $\mh\in[400,500]\gev$. This
is in clear contradiction to the above quoted experimental limits from
the Tevatron for the $[110,200]\gev$ mass range.  Thus, the $WW$
channel already implies that having a light SM-like Higgs boson is
inconsistent with the presence of a 4th generation.  (See also the
earlier analysis of~\cite{Aaltonen:2010sv} using less integrated
luminosity.) The only escape would be if the Higgs boson has
non-standard decays that deplete $\br(h\to WW)$ and $\br(h\to
\gam\gam)$. Since models of this type abound~\cite{cpnsh}, a definitive
conclusion will require actual observation of a Higgs with the
couplings and decays predicted in the SM.

Before leaving the SM, we note from Fig.~\ref{htohsm4} that inclusion
of a heavy sequential $W'$ without a 4th generation gives
$\rgamgam\sim 4-5$ for $\mhsm \lsim 115\gev$, a value that can
probably be excluded relatively soon. But, once $\mhsm\gsim 2m_W$
$\rgamgam$ falls to $\sim 3$, a value requiring large $L$ to either
observe or exclude given that $\Gamma_{\rm tot}^{\hsm}$ is large for
such masses.  If both a 4th generation and a sequential $W'$ are
present the predicted $\rgamgam\sim 15-20$ is probably already
excluded for $\mhsm \lsim 150\gev$ (perhaps higher once the analysis is
done) using the current data set. In contrast, $\rww$ is nearly
unaffected by a possible $W'$.

Even more enhanced signals from the Higgs bosons of
the 2HDM are very possible.
In the context of the 2HDM (a convenient summary appears in the
HHG~\cite{hhg}), the masses of the light and heavy CP-even Higgs
bosons, $h$ and $H$, of the CP-odd Higgs boson, $A$, and
of the charged Higgs boson $H^\pm$ as well as the value of $\tanb$
(the ratio of VEVs for the two doublets) and the CP-even Higgs sector
mixing angle $\alpha$ can all be taken as independent parameters,
whose values will determine the $\lam_i$ of the general 2HDM Higgs
potential. Thus, it is appropriate to present results for each neutral
Higgs boson as a function of its mass for various $\tanb$ values. 

As reviewed in~\cite{hhg}, in the 2HDM there are only two possible
models for the fermion couplings that naturally avoid flavor-changing
neutral currents (FCNC), Model I and Model II. As a brief reminder, we
provide the summary of Table~\ref{coupsummary} of the couplings of the
$h$, $H$ and $A$ in the two cases, relative to SM normalization.
\begin{table}
\caption{Summary of 2HDM quark couplings in Model I and Model II.}
\vspace*{-.2in}
\beq
\begin{array} {|c|c|c|c|c|c|c|}
\hline
\ & \multicolumn{3}{|c|}{\mbox{Model I}} &
 \multicolumn{3}{c|}{\mbox{Model II}}\cr 
\hline
\ & h & H &A & h & H & A \cr
\hline 
t\anti t & {\cosa\over\sinb} & {\sina\over\sinb} & -i\gam_5\cotb  &
{\cosa\over\sinb} & {\sina\over\sinb} & -i\gam_5 \cotb  \cr
\hline
b\anti b & {\cosa\over\sinb} & {\sina\over\sinb} & i\gam_5 \cotb &
-{\sina\over\cosb} & {\cosa\over\cosb} & -i\gam_5\tanb \cr
\hline
\end{array}\nonumber
\label{coupsummary}
\vspace*{-.3in}
\eeq 
\end{table}
In both Model I and Model II the $WW,ZZ$ couplings of the $h$ and
$H$ are given by $\sin(\beta-\alpha)$ and $\cos(\beta-\alpha)$, respectively,
relative to the SM values. And, very importantly, there is no
coupling of the $A$ to $WW,ZZ$ at tree level. 
 If the $\lam_i$ of the Higgs potential are
kept fully perturbative, the decoupling limit, in which $m_H\to m_A$ and
$\sin^2(\beta-\alpha)\to 1$, sets in fairly quickly as $m_A$ increases

\begin{figure}
\vspace*{-.5in}
\includegraphics[height=0.5\textwidth,angle=90]{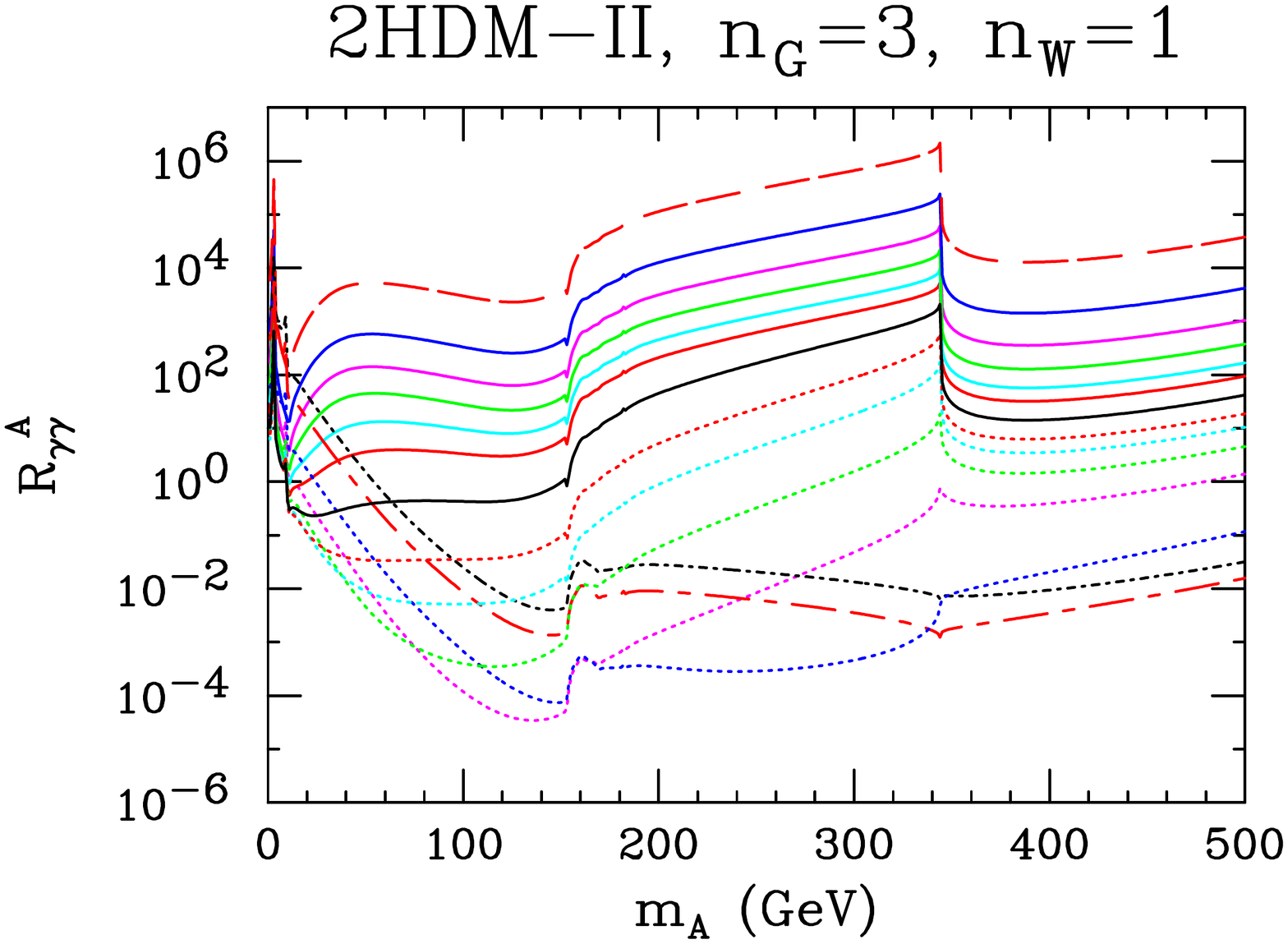}
\vspace*{-.5in}\caption{$\rgamgam$ for the 2HDM-II $A$. The legend is as follows: solid
  black$\to\tanb=1$; red dots$\to\tanb=1.5$; solid
  red$\to\tanb=1/1.5$; cyan dots$\to\tanb=2$; solid
  cyan$\to\tanb=1/2$; green dots$\to\tanb=3$; solid
  green$\to\tanb=1/3$; magenta dots$\to\tanb=1/5$; solid
  magenta$\to\tanb=5$;blue dots$\to\tanb=10$; solid
  blue$\to\tanb=1/10$; long red dashes plus dots$\to\tanb=30$; pure
  long red dashes$\to\tanb=1/30$; black dotdash$\to\tanb=50$. This and
  subsequent 
  figures must be viewed in color in order to resolve the different
  $\tanb$ cases.}
\label{hAII}
\end{figure}

In this Letter, we focus on the 2HDM-II coupling
possibility, and the CP-odd $A$, for which only $\gam\gam$ decays
are relevant.  $\rgamgama$ is plotted as a
function of $m_A$ in Fig.~\ref{hAII} for the 3 generation case.
Enhanced $\gam\gam$ signals, $\rgamgama>1$, are only possible for low
$\tanb$ values.  Although not shown,
enhanced signals are possible for $\tanb<1$ also in Model I.  Note
that $\rgamgama$ is not influenced by possible 
sequential $W'$s since they do not couple to the $A$.

\begin{figure}
\vspace*{-.4in}
\includegraphics[height=0.5\textwidth,angle=90]{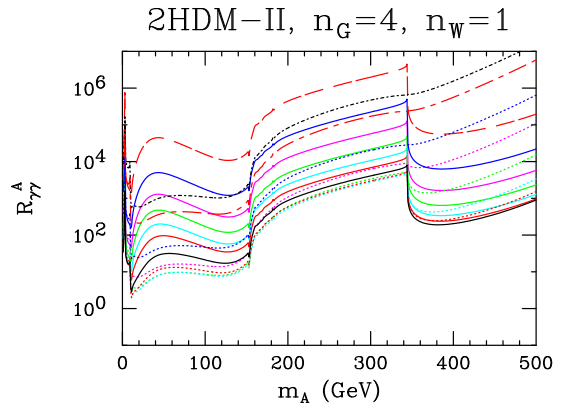}
\vspace*{-.5in}
\caption{$\rgamgam$ for the 2HDM-II $A$ after inclusion of 4th generation
  loops in $gg$ production and in $A\to\gam\gam$ decays. The legend is
  as in Fig.~\ref{hAII}.}
\label{hAII4}
\end{figure}

The impact of a fourth generation on the two-doublet results depends
strongly on whether or not the model is Model~I or Model~II.  In
particular, a 4th generation does not affect $\rgamgama$ in the case of
Model-I.  This is because the $t'$ and $b'$ of the 4th generation
couple to the $A$ with opposite signs but equal coefficients --- see
Table~\ref{coupsummary}.  In contrast, the results for a Model-II $A$
are changed dramatically: the 4th family case is illustrated in
Fig.~\ref{hAII4}.  Regardless of $\tanb$, one predicts large
$\rgamgama$, the smallest values occurring at low $m_A$ for moderate
$\tanb\in[1,5]$, for which $\rgamgama\sim 10$ for
$m_A\in[30,150]\gev$. Of course, this is precisely the range of
$\tanb$ that is preferred in order that the Yukawa coupling of the
$t'$ is perturbative.  $\rgamgama$ increases dramatically for
$m_A>2m_W$ because of the drop in $\br(\hsm\to\gam\gam)$. The enhanced
values of $\rgamgama$ are least likely to be depleted by $A$ decays to
non-SM final states, most particularly $A\to hZ,H^\pm W^\mp$, when $m_A$ is
not large.

%

\begin{figure}
\vspace*{-.4in}
\includegraphics[height=0.5\textwidth,angle=90]{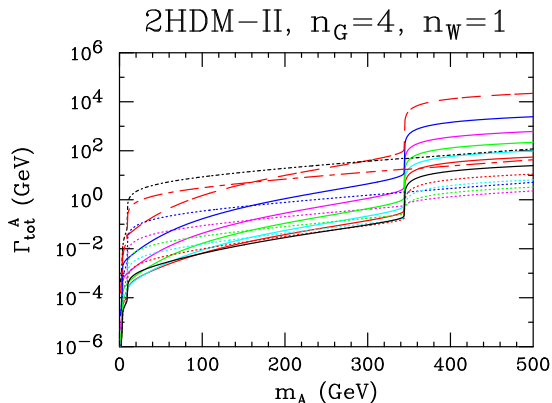}
\vspace*{-.5in}
\caption{$\Gamma_{\rm tot}^A$ for Model II after inclusion of 4th
  generation loops for $A\to gg,\gam\gam$ decays. The legend is as in Fig.~\ref{hAII}.}
\label{hAIIwidth4}
\end{figure}

As noted earlier, a rough estimate using the latest ATLAS plot
shown in~\cite{atlaspheno} suggests
$\rgamgam\lsim 10$ for $\mgamgam\leq 150\gev$.  This estimate assumes
a narrow resonance.  A plot of $\Gamma_{\rm tot}^A$ for $m_A\leq
500\gev$ is given as Fig.~\ref{hAIIwidth4} for the 4 generation case.
Since the $t'$ and $b'$ masses are larger than $m_A/2$, direct decays
to 4th generation quarks do not occur, but the 4th generation quarks
do influence the loop-induced decays to $gg$ (and $\gam\gam$).  For
$m_A<150\gev$, the narrow width approximation only breaks down for
$\tanb\geq 30$.  At $m_A=150\gev$, $\Gamma_{\rm tot}^A=5\gev,13\gev$
for $\tanb=30,50$, respectively.  For such total widths, limits would
then be weaker than naively estimated using the narrow resonance
assumption.  However, we should note that $\tanb>30$ is excluded by
LHC data for $m_A\lsim 170\gev$~\footnote{This assumes the $A$ and $H$
  are not degenerate.} using the $A\to\tauptaum$ decay mode and just
$L=35\pbi$ of data~\cite{Chatrchyan:2011nx}.  These limits will
improve very rapidly with increased $L$.  Once $m_A>2m_t$ the $A$
total width increases dramatically; a study of the feasibility of
detecting a highly enhanced broad $\gam\gam$ signal above the
continuum $\gam\gam$ background is needed to determine the level of
sensitivity.

In passing, we note that $\rgamgamh$ and $\rgamgamH$ for the CP-even
Higgs bosons are less robust as indicators of a 4th generation --- in
particular, they depend significantly on $\sin^2(\beta-\alpha)$ and
are often below 1 (especially for the Yukawa-perturbativity-preferred
modest $\tanb$ values). 
However, it is important to note the complementary of $\rwwh$ and
$\rgamgama$ in the decoupling limit of $\sin^2(\beta-\alpha)=1$.  In
this limit, it is $\rwwh$ that currently does and $\rgamgama$ that
shortly could rule out a 4th generation scenario if the $h$ is
relatively light and if the $A$ is not too heavy, respectively.

Many possible scenarios at the LHC can be envisioned.  For example,
as $L$ increases it could be that a light $A$ ($m_A<200\gev$) is
observed in the $\tauptaum$ mode with rate corresponding to a modest
$\tanb$ value (presumably below $30$ given current limits). If there
is no sign of a $\gam\gam$ peak for the given $L$ it could easily
happen that the limit on $\rgamgam$ will exclude a 4th family in the
Model~II context.  If, on the other hand, no $A$
is detected in the $\tauptaum$ mode a limit on $\tanb$ significantly
below $30$ in the $m_A<200\gev$ mass region is likely.  In this case,
we could only conclude that there can be no 4th generation if we
assume the 2HDM Model II structure and that $m_A<200\gev$; but, of
course, no contradiction would arise if $m_A$ is significantly larger
or if the 2HDM Model II is not the right model.

\begin{figure}
\vspace*{-.4in}
\includegraphics[height=0.5\textwidth,angle=90]{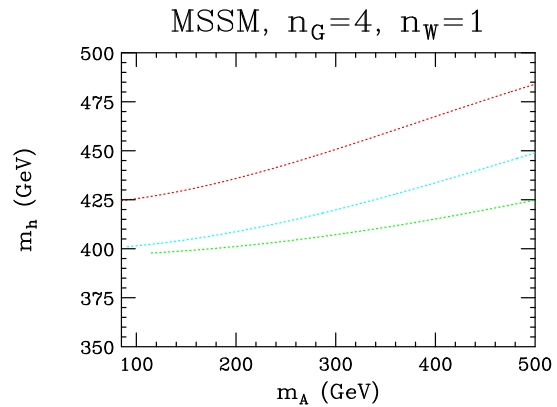}
\vspace*{-.5in}
\caption{$m_h$ vs. $m_A$ for $\tanb=1.5$, $2$ and $3$ --- legend as in
Fig.~\ref{hAII}.}
\label{mhvsma41}
\end{figure}

Let us now focus on the MSSM.  There are many studies of the impact of
a 4th generation on MSSM Higgs
physics~\cite{Cotta:2011bu,Dawson:2010jx,Litsey:2009rp}.
Substantially larger masses than the $400\gev$ value we employ here
are strongly disfavored by precision electroweak constraints and
FCNC considerations~\cite{Dawson:2010jx}. For masses $\sim 400\gev$,
large 4th family loop corrections imply a large mass for the $h$ while
perturbativity for the 4th generation Yukawas requires $\tanb<2-3$.
As stated earlier, for given soft-SUSY-breaking parameters, all Higgs
masses and branching ratios are fixed once $m_A$ and $\tanb$ are
specified. For this study, we employed an extended version of
HDECAY3.60~\cite{Djouadi:1997yw} with ``default'' {\tt hdecay.in}
soft-SUSY-breaking inputs --- 4th generation soft parameters are taken
to be identical to those for the 3rd generation. The resulting values
of $m_h$ as a function of $m_A$ are plotted in Fig.~\ref{mhvsma41}.

\begin{figure}
\vspace*{-.4in}
\includegraphics[height=0.5\textwidth,angle=90]{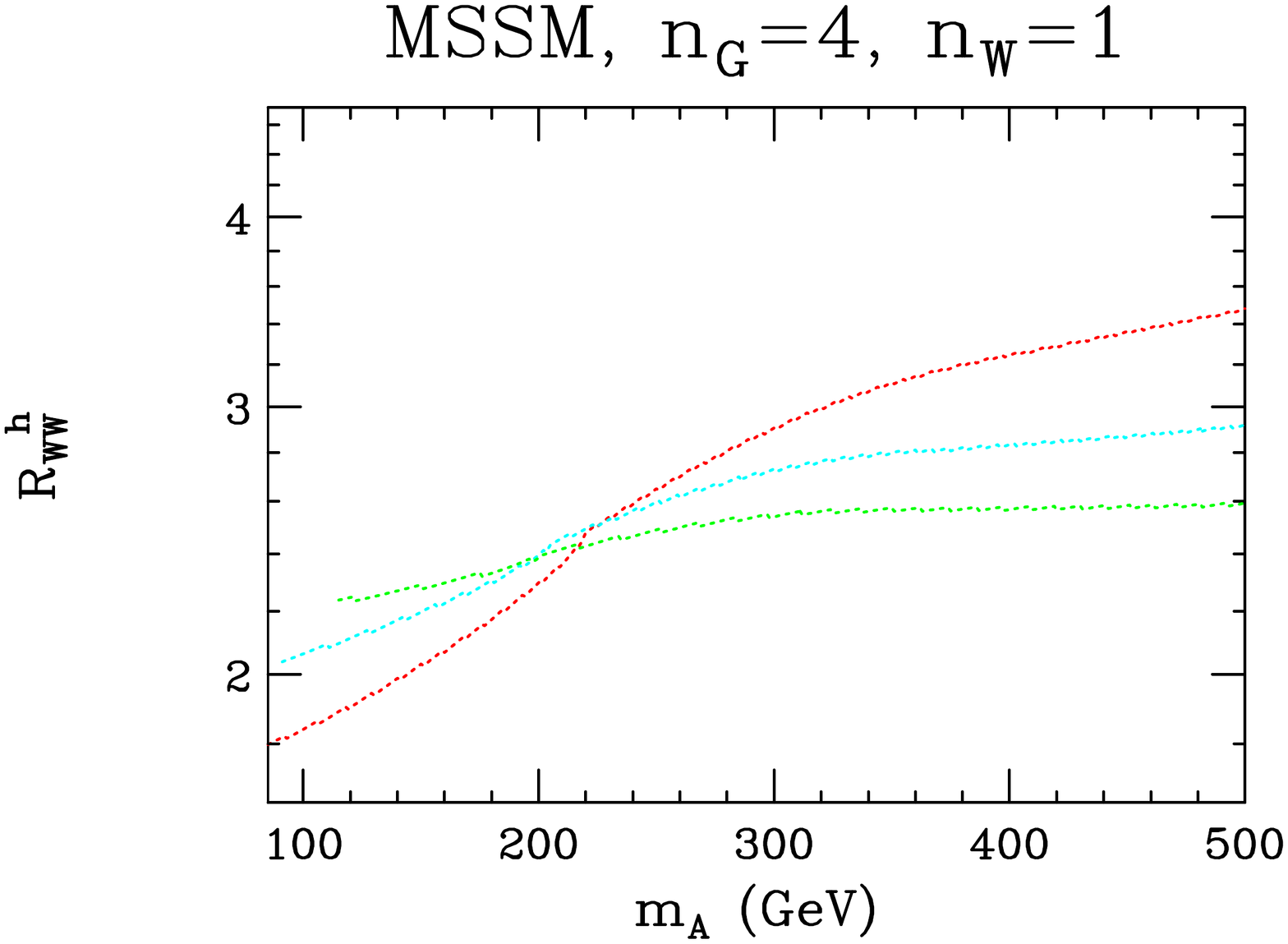}
\vskip -.6in
\includegraphics[height=0.5\textwidth,angle=90]{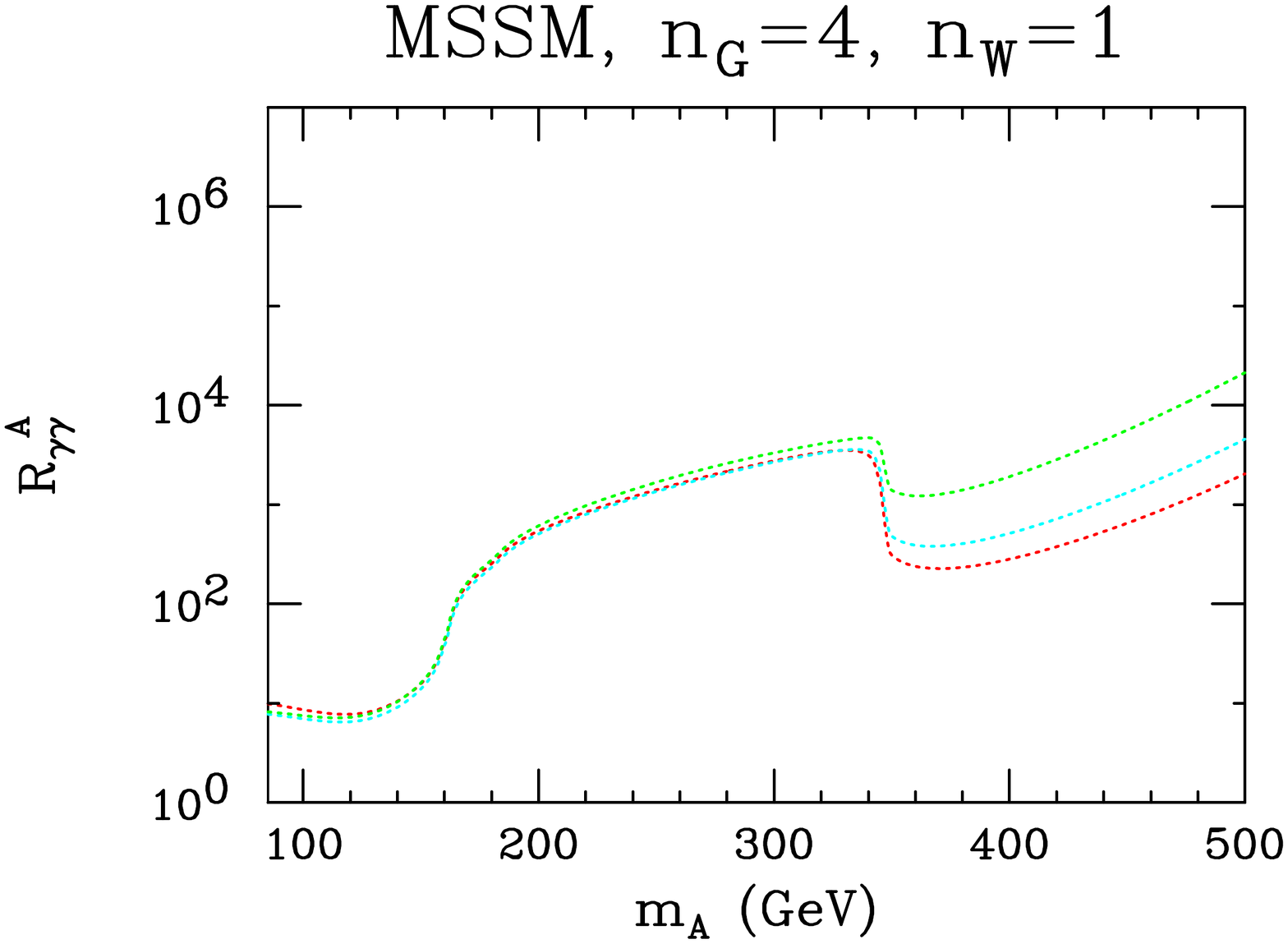}
\vspace*{-.5in}
\caption{MSSM plots for $\tanb=1.5$, $2$ and $3$  --- legend as in
  Fig.~\ref{hAII}. Top: $\rwwh$ vs. $m_A$.  Bottom:
  $\rgamgama$ vs. $m_A$. }
\label{mssmagamgam41}
\end{figure}

Once again, strong constraints on the possible presence of the 4th generation
arise from considering $\rwwh$ and $\rgamgama$.  The relevant plots appear in
Fig.~\ref{mssmagamgam41}.  These plots include loop
effects from both the fermions and the sfermions of the 4th
generation, but the sfermion and other parameters of the default {\tt
  hdecay.in} are such that all sparticles are heavy and do not
contribute significantly to the $h$ or $A$ decays for $m_A<500\gev$.
The smallest values of $\rgamgama\sim 6.5$ occur in
the $m_A<2m_W$ region.  (Ref.~\cite{Cotta:2011bu} also finds
enhancement in $\rgamgama$ at low $m_A$.)  And, as for the 2HDM-II,
for $m_A>2m_W$ one finds $\rgamgama\geq 100$!  $\rwwh$ is
complementary in that for $m_A>200\gev$, $\rwwh>2.4$, a value that
will be probed even at the large $m_{WW}\sim m_h$ values of
Fig.~\ref{mhvsma41} given large enough $L$ at the LHC.  

Thus, we have the following situation. Analysis of LHC $\gamma\gamma$
spectrum data will probably soon place a limit of $\rgamgama<6.5$ out
to $m_A=\mgamgam \sim 2m_W$, in which case a 4th generation will be
inconsistent with the MSSM for $m_A\lsim 2m_W$, barring significant
$A\to SUSY$ decays. For $2m_W<m_A<200\gev$
it seems likely that a limit below the minimum predicted value of
$\rgamgama=100$ will be achieved. Meanwhile, for 4 generations
$\rwwh>2.4$ is predicted for all $m_A\geq 200\gev$ and will eventually be
excludable in the relevant $m_h\sim 400-500 \gev$ mass range.  If
sparticles are light, then hopefully the LHC will detect them and
$\rwwh$ and $\rgamgama$ predictions can be corrected for substantial
$\br(h,A\to SUSY)$ values.  In addition, predictions for
$\Gamma_{gg}^{h,A}\br(h,A\to SUSY)$ will be larger in the presence of
a 4th generation than without.
 
Finally, we note that if there is a $W'$, $\rgamgama$ is not affected
(because of the absence of a tree-level $AW'W'$ coupling) while
changes to $\rwwh$ are very tiny. Further, $\rwwh$ is only modestly
influenced by sfermion loop contributions to $\Gamma_{gg}^h$ and sfermion
loops are not present for either $gg\to A$ or $A\to \gam\gam$. Thus,
$\rgamgama$ and $\rwwh$ are quite robust tests for the presence of a 4th
generation and can potentially eliminate the possibility of 4
generations in the context of the MSSM even if no Higgs is observed.
Of course, by the time sufficient $L$ is available to measure $\rwwh$
out to large $m_h$, direct observation or exclusion of the
4th-generation quarks may have occurred.

Once a $\gam\gam$ or $WW$ peak emerges (as will eventually happen if
there is one ore more light Higgs bosons) a multitude of possibilities
will need to be analyzed. If no Higgs has been seen in any other mode,
then there will be a plethora of Higgs sector choices that could
explain the $\gam\gam$ or $WW$ peak, both in the general 2HDM context and in
the MSSM.  In the MSSM context, if $\tanb$ is known from general observations of
superpartners, it will be important to see if there is a Higgs boson
within some Higgs scenario that can explain the peak for the known
$\tanb$ value, either with or without a 4th generation and/or $W'$.

To summarize, we have shown that great importance attaches to the most
exhaustive possible search for peaks and enhancements in the
$\gam\gam$, $WW$ and $ZZ$ mass spectra over the broadest possible
range of $\mgamgam$, $m_{WW}$, and $m_{ZZ}$.  Either detection of a
peak or a simple limit on $\rgamgam$, $\rww$ and $\rzz$ as a function
of $\mgamgam$, $M_{WW}$, and $M_{ZZ}$ will provide highly significant
constraints and/or consistency checks both on the Higgs sector and on
the possible existence of a 4th generation or $W'$.

\vspace*{-.2in}
\acknowledgments 

JFG is supported by U.S. DOE grant No. DE-FG03-91ER40674. Thanks to
P.~Jaiswal for noting the incorrect results for the MSSM $h$ in the
first version of the paper.

\end{document}